\documentclass[aps,prl,reprint,superscriptaddress]{revtex4-1}

\usepackage{graphicx}
\usepackage{dcolumn}
\usepackage{bm}
\usepackage{amsfonts}
\usepackage{amssymb}
\usepackage{amsmath}
\usepackage{graphicx}
\usepackage{color}
\usepackage{array}
\usepackage{dcolumn} 
\usepackage{bm} 
\usepackage{amsthm}
\usepackage{latexsym}
\usepackage{float}
\usepackage{tabularx}

\begin{document}


\title{Thermo-Optical Chaos and Direct Soliton Generation in Microresonators}


\author{Chengying Bao}
\email{bao33@purdue.edu}
\affiliation{School of Electrical and Computer Engineering, Purdue University, 465 Northwestern Avenue, West Lafayette, IN 47907-2035, USA}

\author{Yi Xuan}
\affiliation{School of Electrical and Computer Engineering, Purdue University, 465 Northwestern Avenue, West Lafayette, IN 47907-2035, USA}
\affiliation{Birck Nanotechnology Center, Purdue University, 1205 West State Street, West Lafayette, Indiana 47907, USA}

\author{Jose A. Jaramillo-Villegas}
\affiliation{School of Electrical and Computer Engineering, Purdue University, 465 Northwestern Avenue, West Lafayette, IN 47907-2035, USA}
\affiliation{Facultad de Ingenier\'{i}as, Universidad Tecnol\'{o}gica de Pereira, Pereira, RI 66003, Colombia}

\author{Daniel E. Leaird}
\affiliation{School of Electrical and Computer Engineering, Purdue University, 465 Northwestern Avenue, West Lafayette, IN 47907-2035, USA}

\author{Minghao Qi}
\affiliation{School of Electrical and Computer Engineering, Purdue University, 465 Northwestern Avenue, West Lafayette, IN 47907-2035, USA}
\affiliation{Birck Nanotechnology Center, Purdue University, 1205 West State Street, West Lafayette, Indiana 47907, USA}

\author{Andrew M. Weiner}
\email{amw@purdue.edu}
\affiliation{School of Electrical and Computer Engineering, Purdue University, 465 Northwestern Avenue, West Lafayette, IN 47907-2035, USA}
\affiliation{Birck Nanotechnology Center, Purdue University, 1205 West State Street, West Lafayette, Indiana 47907, USA}
\begin{abstract}
We investigate, numerically and experimentally, the effect of thermo-optical (TO) chaos on direct soliton generation (DSG) in microresonators. When the pump laser is scanned from blue to red and then stopped at a fixed wavelength, we find that the solitons generated sometimes remain (survive) and sometimes annihilate subsequent to the end of the scan.  We refer to the possibility of these different outcomes arising under identical laser scan conditions as coexistence of soliton annihilation and survival.  Numerical simulations that include the thermal dynamics show that the coexistence of soliton annihilation and survival is explained by TO chaos accompanying comb generation.  The random fluctuations of the cavity resonance occurring under TO chaos are also found to trigger spontaneous soliton generation after the laser scan ends. The coexistence of soliton annihilation and survival as well as spontaneous soliton generation are observed experimentally in a silicon-nitride microresonator. The elucidation of the role of TO chaos provides important insights into the soliton generation dynamics in microresonators, which may eventually facilitate straightforward soliton generation in fully-integrated microresonators.
\end{abstract}


\maketitle
Chaos is a fundamental phenomenon that widely exists in physics, biology, engineering \cite{Strogatz2014nonlinear} as well as  in financial market \cite{Hsieh1991chaos}. In optical microreosnators, the interplay between optical field and mechanical motion can lead to chaos \cite{Vahala_PRL2007chaotic,Yang_NP2016optomechanically}. However, the interplay between the optical field and thermal field generally does not lead to chaos in weakly pumped microresonators, due to thermal locking effect \cite{Vahala_OE2004dynamical}. In strongly pumped microresonators, nonlinearity can lead to Kerr comb generation \cite{Kippenberg_Nature2007optical,Kippenberg_Science2011microresonator}. Some states of Kerr combs exhibit chaos in the intracavity power under dynamics governed by nonlinearity and dissipation \cite{Naoaki_JPSJ1985chaotic,Matsko_OL2013chaotic,Kippenberg_NP2014temporal,Weiner_NP2015mode,Chembo_Chaos2014routes,
Coen_Optica2016observations}. This optical chaos can result in thermal chaos via the thermo-optical (TO) effect, which shifts the cavity resonance and the pump frequency detuning stochastically \cite{Vahala_OE2004dynamical}. Since the comb properties, including coherence, bandwidth, output power \cite{Kippenberg_NP2014temporal,Kippenberg_Science2016photonic,Coen_OL2013universal,Bao_PRA2015carrier,
Kippenberg_NP2012universal,Weiner_PRL2016observation,Coen_OE2013dynamics} are sensitive to pump frequency detuning, thermal chaos will act back on the optical chaos. We refer this interplay between optical chaos and thermal chaos as TO chaos.

The influence of TO chaos is especially important for the soliton Kerr comb generation \cite{Kippenberg_NP2014temporal,Kippenberg_Science2016photonic}, as the transition from the chaotic state to the soliton state is accompanied by a significant drop in the intracavity power, which strongly blue-shifts the cavity resonance. This blue-shift could cause the pump to fall out of resonance. Hence, the soliton generation exhibits a transient instability in microresonators with strong thermal effect, such as silicon-nitride (Si$_3$N$_4$) \cite{Kippenberg_Science2016photonic,Weiner_OE2016intracavity,Gaeta_OL2016thermally,Kippenberg_PRL2016raman} and silica (SiO$_2$) cavities \cite{Vahala_Optica2015soliton}. To mitigate the transient instability, complicated experimental methods are needed. For instance,``power-kicking" \cite{Kippenberg_Science2016photonic,Kippenberg_arXiv2016bringing,Vahala_OL2016active,Coen_OL2016experimental} or backward tuning \cite{Weiner_OE2016intracavity,Gaeta_OL2016thermally,Kippenberg_NP2016universal,
Gaeta_arXiv2016chip}, are usually needed for soliton generation in Si$_3$N$_4$ or SiO$_2$ cavities. The``power-kicking" method needs additional devices (acousto-optic modulators or electro-optic modulators). The backward tuning method is limited by the response time of lasers or heaters and only works for samples exhibiting soliton-step of long duration. For full integration of the comb system and out-of-lab applications, it is better to avoid the additional devices and complicated experimental methods. Hence, direct soliton generation (DSG), which we define as simply tuning the laser from the blue to the red and stopping at a specific wavelength to excite the soliton, is highly desired. DSG was first demonstrated in MgF$_2$ whispering gallery mode cavities \cite{Kippenberg_NP2014temporal}, and it benefits from the low TO coefficient of MgF$_2$ (an order of magnitude lower than Si$_3$N$_4$ and SiO$_2$) \cite{Kippenberg_NC2013mid}. DSG is also possible in Si microresonators, but assistance from multi-photon absorption and free carrier absorption are needed \cite{Gaeta_Optica2016}. DSG has also been demonstrated in chip-scale Si$_3$N$_4$ microresonators \cite{Kippenberg_Optica2016photonic,Diddams_arXiv2016stably}; however, the generation dynamics remains largely unexplained and the role of TO chaos is not unveiled. Since Si$_3$N$_4$ has low loss in the telecom band and is compatible with CMOS integration  \cite{Gaeta_NP2010cmos,Lipson_NP2013new}, understanding of DSG in Si$_3$N$_4$ microresonators is critical for the integration of soliton Kerr combs, which could greatly shrink the footprint of light sources for optical communications \cite{Kippenberg_NP2014coherent,Kippenberg_arXiv2016microresonator,Weiner_CLEO2016long}, spectroscopy \cite{Gaeta_arXiv2016chip,Vahala_Science2016microresonator,Gaeta_arXiv2016silicon}, arbitrary optical waveform generation \cite{Weiner_NP2011spectral} etc.

\begin{figure*}[t]
\centering
\includegraphics[width=0.82\linewidth]{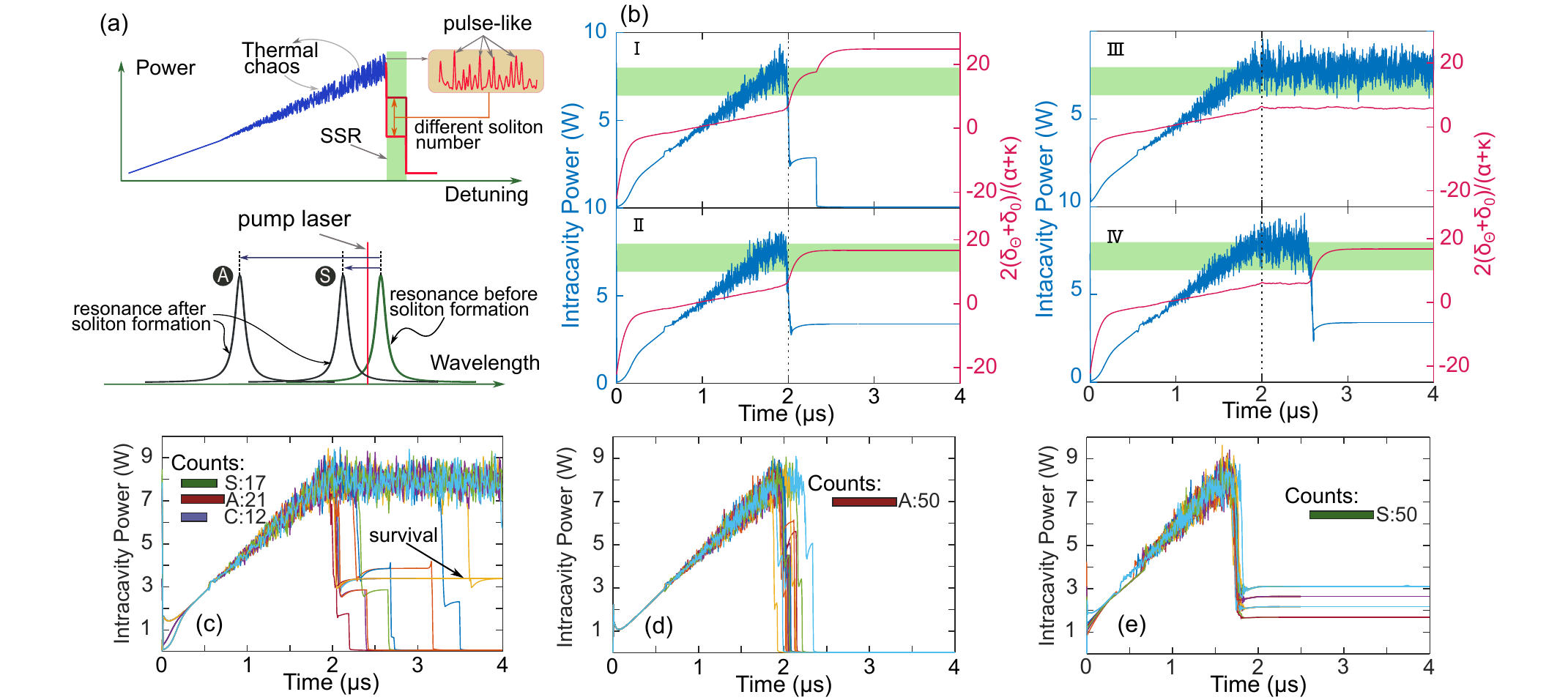}
\caption{(color online) (a) Illustration of the thermo-optical (TO) chaos and its effect on direct soliton generation (DSG). How closely the pulse-like structures in the chaotic waveform (shown in the shaded box) are to the sech shape of soliton can affect how many solitons are generated and the power drop upon soliton formation. Depending on the thermal shift associated with the power drop in the soliton generation, the system can exhibit the coexistence of soliton survival and soliton annihilation. SSR: soliton supporting range. (b) Various routes to soliton generation in DSG: solitons can form around the end the laser scan and annihilate (\uppercase\expandafter{\romannumeral1}) or survive (\uppercase\expandafter{\romannumeral2}). When solitons survive, $\Delta_\text{{eff}}$ stays in the SSR (illustrated by the green box), which is mapped to be $9.2<\Delta_\text{{eff}}<17.9$ with thermal effect turned off. The comb can remain chaotic when the laser scan ends (\uppercase\expandafter{\romannumeral3}, \uppercase\expandafter{\romannumeral4}). When the laser frequency is held constant, the comb can remain in the chaotic regime (\uppercase\expandafter{\romannumeral3}) or form solitons spontaneously (\uppercase\expandafter{\romannumeral4}). Power change in 50 scans (only 20 power traces are plotted for clarity) under (c) modest thermal effects, $\xi=-4.5\times10^{4}$ (Ws)$^{-1}$; note that the results for several distinct trials with soliton survival overlap closely and are hard to distinguish. (d) strong thermal effects $\xi=-1.2\times10^{5}$ (Ws)$^{-1}$, and (e) weak thermal effects, $\xi=-1.2\times10^{4}$ (Ws)$^{-1}$. The coexistence of soliton annihilation and survival happens for the modest thermal effects. S: survival, A: annihilation, C: chaotic.}
\label{Fig1Simulation}
\end{figure*}

In this Letter, we investigate DSG in Si$_3$N$_4$ microresonators numerically and experimentally. In addition, we show two phenomena in DSG: 1) solitons generated during identical laser scans in different experimental trials can either remain (survive) or annihilate subsequent to the end of the scan (we refer to this as ``coexistence of soliton annihilation and survival”), 2) delayed spontaneous soliton generation after the end of the laser scan. Although number of the generated solitons has been shown to be stochastics \cite{Kippenberg_NP2014temporal,Kippenberg_Science2016photonic,Kippenberg_NP2016universal,Weiner_OE2015deterministic}, the random nature in soliton annihilation and survival due to TO chaos has not been predicted or demonstrated yet, to the best of our knowledge. By modeling the soliton generation in Si$_3$N$_4$ microresonators using the generalized Lugiato-Lefever equation (LLE) \cite{Lugiato_PRL1987spatial,Coen2013modeling}, supplemented by the thermal dynamics \cite{Kippenberg_NP2016universal,Tanabe_PJ2016thermal,Gorodetsky_OE2016harmonization}, we numerically predict the coexistence of soliton annihilation and survival in DSG. The presence of TO chaos also allows various routes to soliton formation, including a spontaneous soliton generation route, where the Kerr comb transits to the soliton state spontaneously after the laser scan ends (while the laser frequency is held constant). In experiments, we demonstrate DSG in a Si$_3$N$_4$ microreosnator and observe the simulated coexistence of soliton annihilation and survival, as well as the spontaneous soliton generation.

In Kerr comb generation, solitons are only supported in a narrow range of detuning, which we refer to as the soliton supporting range (SSR) \cite{Kippenberg_NP2014temporal,Coen_OL2013universal,Weiner_PRL2016observation} (see Fig. 1(a)). Moreover, the characteristics of the waveform in the chaotic state will affect how many pulse-like structures can develop into solitons. The intracavity power drop and the corresponding cavity resonance blue-shift that occur with transition from chaotic to soliton states depend on the power prior to soliton formation and the number of solitons generated and will differ in repeated trials. In some cases, the pump frequency detuning stays within the SSR, resulting in soliton survival in DSG; while in the other cases, detuning is shifted out of the SSR, resulting in soliton annihilation, as illustrated in Fig. 1(a).

To analyze the soliton generation dynamics quantitatively, we start our investigation of the soliton generation with numerical simulation based on the generalized LLE \cite{Lugiato_PRL1987spatial,Coen2013modeling}, including thermal effects \cite{Kippenberg_NP2016universal}. 
In simulations, thermal effects will have an additional pump phase detuning in the LLE. Here, we write the generalized LLE as \cite{Kippenberg_NP2016universal,Weiner_PRL2016observation}

\begin{equation}
\begin{aligned}
& \left( {{t}_{R}}\frac{\partial }{\partial t}+\frac{\alpha +\kappa }{2}+i\frac{{{\beta }_{2}}L}{2}\frac{{{\partial }^{2}}}{\partial {{\tau }^{2}}}+i\left( {{\delta }_{0}}+{{\delta }_{\Theta }} \right) \right)E-\sqrt{\kappa }{{E}_{in}} \\
& -i\gamma \left( 1+\frac{i}{{{\omega }_{0}}} \right)L\left( E\int_{-\infty }^{+\infty }{R\left( {{\tau }'} \right){{\left| E\left( t,\tau -{\tau }' \right) \right|}^{2}}}d{\tau }' \right)=0, \\
\end{aligned}
\label{eq:LLE}
\end{equation}

\begin{equation}
\frac{d{{\delta }_{\Theta }}}{dt}=-\frac{{{\delta }_{\Theta }}}{{{\tau }_{0}}}+\xi P.
\label{eq:relaxation}
\end{equation}
Here, $E$, $t_R$, $L$, $\omega_0$, $\beta_2$, $\gamma$, $\alpha$, $\kappa$ are the envelope of the intracavity field, round-trip time, cavity length, carrier frequency, group-velocity dispersion, nonlinear coefficient, intrinsic and coupling loss, respectively; $R(\tau)$ is the nonlinear response, including the instantaneous electronic and delayed Raman response (see Ref. \cite{Weiner_PRL2016observation} for its calculation method); $\delta_0=(\omega_{r0}-\omega_p)t_R$ is the laser detuning ($\omega_{r0}$ is the cold cavity resonance, $\omega_{p}$ is the pump frequency), $\delta_\Theta=(\omega_{r\Theta}-\omega_{r0})t_R$ is the thermal detuning ($\omega_{r\Theta}$ is the cavity resonance shifted by thermal effects, but not by the Kerr effect). The thermal dynamics is included in Eq. \ref{eq:relaxation}, where $P$ is the average intracavity power, and $\xi$ is the coefficient describing the shift of the detuning in response to the average intracavity power, which depends on the material absorption and the fraction of power absorbed converted to heat and is chosen as $-4.5\times10^{4}$ (Ws)$^{-1}$ in simulations; $\tau_0$ is the thermal response time, which is measured to be sub-microsecond in our fabricated microresonators previously \cite{Weiner_CLEO2014fast}. Note that $\tau_0$ also depends on the design of the cavity and we set it to be 100 ns to reduce simulation time. To resemble common resonators, we choose parameters as: $\alpha+\kappa$ =0.0037, $\kappa|E_{in}|^2$=0.11 mW, $\beta_2=-61$ ps$^2$/km, $\gamma$ =1.4 (Wm)$^{-1}$, $L$=628 $\mu$m. For brevity, we denote and normalize the laser detuning, thermal detuning, and effective detuning as, $\Delta_0=2\delta_0/(\alpha+\kappa)$, $\Delta_\Theta=2\delta_\Theta/(\alpha+\kappa)$ and $\Delta_{\text{eff}}=2(\delta_0+\delta_\Theta)/(\alpha+\kappa)$, respectively.

The simulation starts from noise and 
we tune $\Delta_0$ linearly from $-$3 to 25 in 0$\sim$2 $\mu$s and hold it at 25 in 2$\sim$4 $\mu$s to trigger soliton formation. For the chosen $\xi$, various soliton generation dynamics exist under the same scan conditions. In Fig. \ref{Fig1Simulation}(b)\uppercase\expandafter{\romannumeral1}, the Kerr comb starts to form into solitons around the end of the laser tuning. With the power drop in the soliton transition, thermal detuning increases (becomes less negative) and $\Delta_{\text{eff}}$ increases, exceeding the SSR, which is indicated by the green shaded boxes in Fig. \ref{Fig1Simulation}(b) (SSR for the used pump power is $9.2<\Delta_{\text{eff}} <17.9$, which is mapped out by scanning $\Delta_0$ from $-$3 to 25 without thermal effects). Consequently, solitons annihilate due to thermal induced resonance blue-shift. In contrast, in some other scans $\Delta_{\text{eff}}$ remains in the SSR after the soliton transition, and the solitons survive in DSG (Fig. \ref{Fig1Simulation}(b)\uppercase\expandafter{\romannumeral2}). Furthermore, the presence of TO chaos also allows distinct routes to soliton formation. For instance, in some scans the comb remains in the chaotic state with a blue-detuned pump over the full 4 $\mu$s simulation (Fig. \ref{Fig1Simulation}(b)\uppercase\expandafter{\romannumeral3}). However, the transition to the soliton can happen spontaneously, even the comb remains in a chaotic state for finite time after the laser scan stops (Fig. \ref{Fig1Simulation}(b) \uppercase\expandafter{\romannumeral4}). Similar spontaneous transition to coherent state has also been observed in soliton crystal generation in silica microresonators but remains unexplained \cite{Diddams_CLEO2016stabilizing}. Here, we attribute the spontaneous soliton generation to TO chaos. The intracavity power changes stochastically in the chaotic state; if the power drops to some level ($\Delta_\Theta$ becomes less negative) and $\Delta_{\text{eff}}$ increases to approach the SSR, soliton formation is triggered.

In Fig. \ref{Fig1Simulation}(c), we show the coexistence of soliton annihilation and survival in 50 scans (only 20 scans are shown for clarity). The existence ratio of soliton annihilation and survival is comparable with $\xi=-4.5\times10^{4}$ (Ws)$^{-1}$. The annihilation or survival of solitons is strongly affected by the final intracavity power. For all the cases of soliton survival, the soliton number (6) and the final power are the same.
This is because certain level of final intracavity power is needed to have a final $\Delta_\Theta$ to hold $\Delta_{\text{eff}}$ in the SSR. Thus, the soliton number is constrained. If the soliton number is too large or too small, the final $\Delta_\Theta$ cannot support $\Delta_{\text{eff}}$ in the SSR and solitons annihilate. Since soliton number depends on the waveform of the chaotic state, TO chaos is critical for the soliton annihilation and survival. Therefore, TO chaos constitutes a new mechanism for the coexistence of different states in the same nonlinear system, which is different from the coexistence of different mode-locking states in a fiber laser, arising from the spectral filtering and high order nonlinearity \cite{Cundiff_PRL2015observation}. The coexistence of soliton annihilation and survival only occurs a certain range of the thermal strength parameter. For instance, solitons tend to annihilate in all the scans under stronger thermal effects with $\xi=-1.2\times10^{5}$ (Ws)$^{-1}$ (Fig. \ref{Fig1Simulation}(d)). With larger $\xi$, $\Delta_0$ should end at a larger value to induce soliton transition (tuned from $-$3 to 57 and held at 57), thus requiring a large soliton number (high final power) to hold $\Delta_{\text{eff}}$ in the SSR for soliton survival (see Sec. 1 of Supplementary Information for an estimate of the needed final power). Although the soliton number is stochastic, it follows some distribution  \cite{Kippenberg_NP2016universal,Weiner_OE2015deterministic} and too large a soliton number may not be accessible. For weak thermal effects, e.g., $\xi=-1.2\times10^{4}$ (Ws)$^{-1}$, $\Delta_0$ is tuned from $-$3 to 14 and held at 14 to induce soliton transition and solitons survive in all the 50 scans (Fig. \ref{Fig1Simulation}(e)). Moreover, the soliton number is random, as soliton number will not affect $\Delta_{\text{eff}}$ significantly for weak thermal effects.

\begin{figure}[t]
\centering
\includegraphics[width=0.95\linewidth]{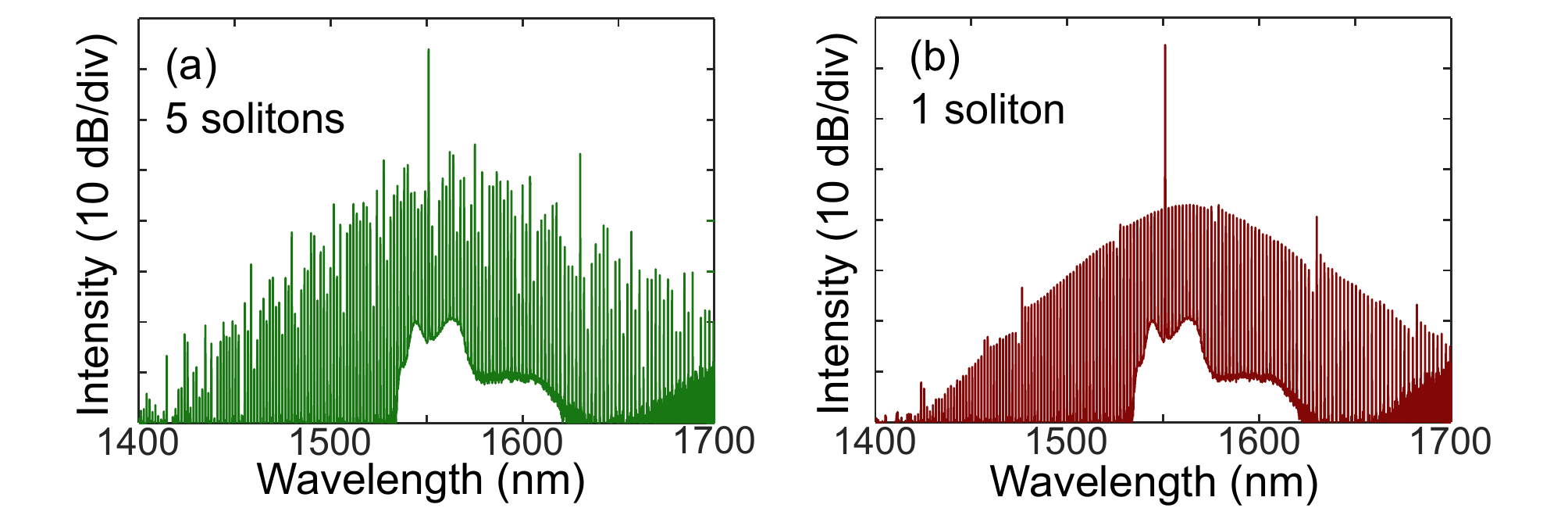}
\caption{(color online) Spectra of the solitons generated in DSG, (a) multi-solitons, (b) single-soliton.}
\label{Fig2Spectrum}
\end{figure}

\begin{figure}[t]
\centering
\includegraphics[width=0.95\linewidth]{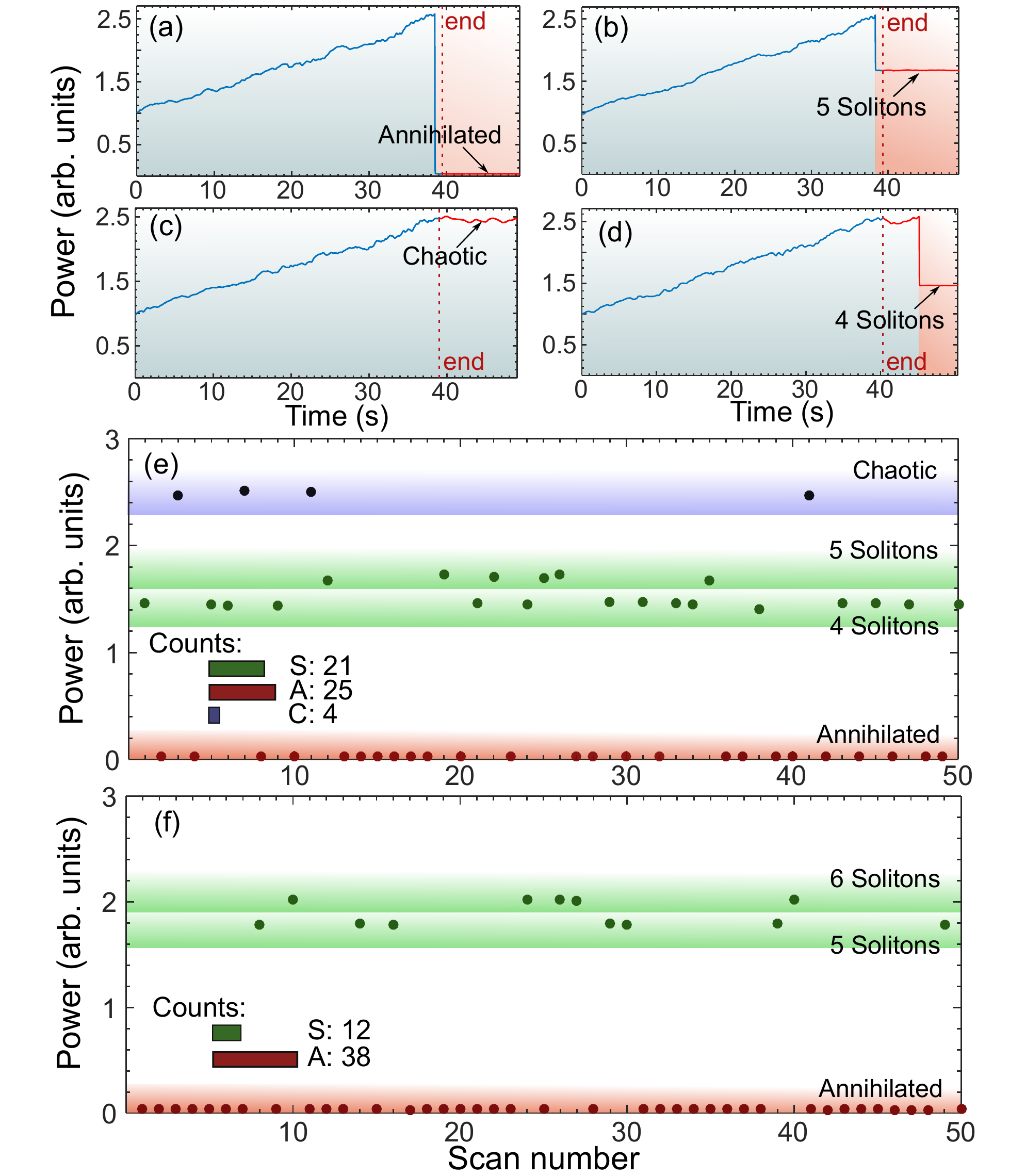}
\caption{(color online) Generation dynamics of DSG: (a) solitons form before laser scan ends and annihilate, (b) soliton forms before laser scan ends and survive, (c) comb remains chaotic, soliton not formed, (d) comb remain chaotic when laser scan ends and soliton forms spontaneously. The dashed lines in (a-d) shows the moments when the laser scan ends; the blue lines are power trace recorded before laser scan ends, while red lines show the power trace after laser scan ends; the blue and red shaded regions shows the laser is effective blue-detuned (chaotic) or red-tuned (soliton formed) relative to the hot cavity resonance. (e) The final intracavity power 10 seconds after the laser scan ends. The power levels show the operation states of the final comb, including chaotic (C), soliton annihilation (A) and survival (S). (f) The final intracavity power when using a higher pump power and larger laser detuning. The soliton number tends to be larger.}
\label{Fig3Scan}
\end{figure}

In experiments, we test DSG in a Si$_3$N$_4$ microreosnator, with a loaded Q-factor of 1.1$\times$10$^6$, geometry of 800$\times$2000 nm, 100 $\mu$m radius, dispersion of $-$60~ps$^2$/km. The tuning of the laser is calibrated by a wavemeter with 20 MHz resolution, well below the linewidth of the pumped resonance (180 MHz). Hence, laser scan starts and stops at the same wavelength for different scans ($\Delta_0$ is tuned in the same way). The converted comb power (removing the pump by a notch filter) is measured during the laser scan (See Sec. 2 of Supplementary Information). Since acquiring data from wavemeter and power-meter takes $\sim$10 ms, tuning of the laser is much slower than in simulations. However, we find the slower tuning still captures the general dynamics of DSG (see Sec. 3 in Supplementary Information, which also shows the thermal response time has a minor effect on DSG but is critical to the soliton-step length).

We first map out the soliton-step of the studied cavity and choose a wavelength around the middle of the step as the stop wavelength (chosen as 1551.0301 nm, Sec. 2 in Supplementary Information) under the pump power of 200 mW. Then, we tune the laser to the stop wavelength and hold at the stop wavelength to demonstrate DSG. Soliton Kerr combs are generated successfully; the cavity generally supports multiple solitons, with a representative spectrum shown in Fig. \ref{Fig2Spectrum}(a). The spectrum is structured with the center frequency shifted to the red of the pump due to Raman effects (soliton-self-frequency shift), which is evidence of soliton operation, as other states do not exhibit this shift \cite{Kippenberg_PRL2016raman}. After the solitons stabilize, we can get the single-soliton state, with smooth sech$^2$-like spectrum (Fig. \ref{Fig2Spectrum}(b)), by backward tuning the laser. By measuring the power change during backward tuning \cite{Kippenberg_NP2016universal}, soliton number can be deduced. The comb in Fig. \ref{Fig2Spectrum}(a) is found to consist of 5 solitons.

We then run 50 laser scans (scan to the stop wavelength and hold for $\sim$10 s) and record the power evolution to verify the coexistence of soliton annihilation and survival and spontaneous soliton generation. In Figs. \ref{Fig3Scan}(a-d), we show various routes to soliton generation in DSG. In some scans, the transition to solitons can happen before the scan ends (Figs. \ref{Fig3Scan}(a), (b)). However, the comb can also be chaotic, when the laser scan stops (Figs. \ref{Fig3Scan}(c), (d)). If the comb is chaotic, the comb may switch to a soliton state spontaneously in some cases after the scan stops (Fig. \ref{Fig3Scan}(d)). From Figs, \ref{Fig3Scan}(a-d), we can also see the coexistence of soliton annihilation and survival. To further demonstrate this coexistence, we show the final power after holding 10 s in Fig. \ref{Fig3Scan}(e), which clearly shows both soliton annihilation and survival states. Furthermore, the fraction of soliton survival and annihilation outcomes are comparable. Moreover, we also see the power restriction in the soliton survival state, as simulations predict; the soliton number is limited to 4 and 5. We also note that DSG is not restricted to a very specific pump power and stopping wavelength. For instance, DSG can be achieved under larger laser detuning (stop at 1551.0333 nm) and higher pump power (300 mW). In that case, the soliton number tends to be larger for soliton survival (Fig. \ref{Fig3Scan})(f). The intracavity power before soliton formation becomes higher with stronger pump. To avoid a strong intracavity power drop upon soliton formation and balance the larger laser detuning to hold $\Delta_{\text{eff}}$ in the SSR, a larger number of solitons is needed. The increase in soliton number is consistent with simulations (see Sec. 4 of the Supplementary Information, which also shows the spontaneous soliton generation under 300 mW pump power).

In conclusion, we show DSG in Si$_3$N$_4$ microresonators numerically and experimentally. Soliton annihilation and survival can coexist in DSG for cavities with modest thermal effects. We also observe various routes to soliton formation, including spontaneous soliton generation from a chaotic state. The spontaneous soliton generation and slow laser tuning speed used in DSG show rapid and precise control of the laser tuning is not necessarily required. 
Lowering thermal effects can facilitate simple DSG. The contribution to thermal absorption in Si$_3$N$_4$ cavities mainly lies in the N-H bond absorption, which can be suppressed by annealing \cite{Habraken1991lpcvd,Adibi_OE2013vertical,Wong_SR2015low}. With more investigation of the mechanism of the absorption contributing to heat in microresonators, simple method can be used for the soliton Kerr combs generation in various materials cavities.

\begin{acknowledgments}
This work was supported in part by the Air Force Office of Scientific Research (AFOSR) (Grant No. FA9550-15-1-0211), by the DARPA PULSE program (Grant No. W31P40-13-1-0018) from AMRDEC, and by the National Science Foundation (NSF) (Grant No. ECCS-1509578).
\end{acknowledgments}



\bibliography{reflist}
\end{document}



\title{Supplementary Information for ``Thermo-Optical Chaos and Direct Soliton Generation in Microresonators"}

\author{Chengying Bao$^1$}
\email{bao33@purdue.edu}


\author{Yi Xuan$^{1,2}$}

\author{Jose A. Jaramillo-Villegas$^{1,3}$}

\author{Daniel E. Leaird$^{1}$}

\author{Minghao Qi$^{1,2}$}

\author{Andrew M. Weiner$^{1,2}$}
\email{amw@purdue.edu}
\affiliation{$^{1}$School of Electrical and Computer Engineering, Purdue University, 465 Northwestern Avenue, West Lafayette, IN 47907-2035, USA}
\affiliation{$^{2}$Birck Nanotechnology Center, Purdue University, 1205 West State Street, West Lafayette, Indiana 47907, USA}
\affiliation{$^{3}$Facultad de Ingen\'{i}eras, Universidad Tecnol\'{o}gica de Pereira, Pereira, RI 66003, Colombia}

\maketitle

\renewcommand{\thefigure}{S\arabic{figure}}
\renewcommand{\theequation}{S\arabic{equation}}

\renewcommand*{\citenumfont}[1]{S#1}
\renewcommand*{\bibnumfmt}[1]{[S#1]}
\renewcommand{\thesection}{\arabic{section}}
\section{Estimation of final power}
\begin{figure}[h]
\includegraphics[width=0.9\columnwidth]{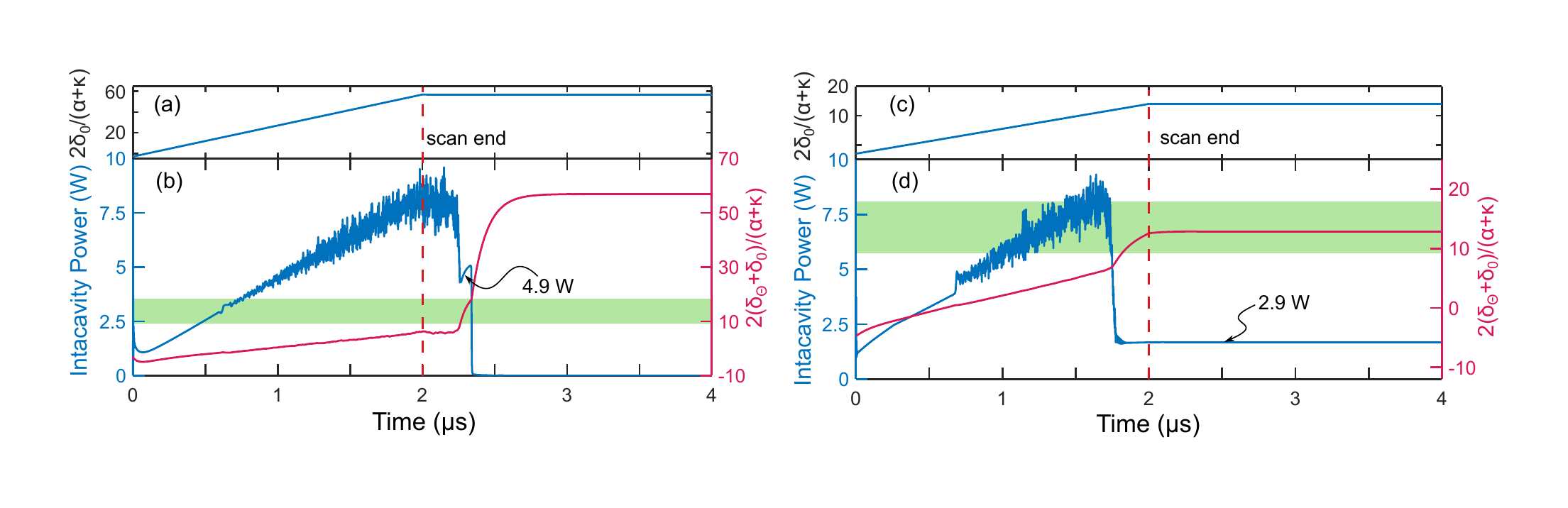}
\caption{(color online)  (a) Laser tuning method for soliton generation under strong thermal effects, (b) the corresponding intracavity power and effective detuning dynamics, showing soliton annihilation. (c) Tuning of the laser for soliton generation under weak thermal effects, (b) the corresponding intracavity power and effective detuning dynamics, showing soliton survival. The green box shows the soliton supporting range (SSR).}
\label{FigS1Strong}
\end{figure}

The final power is closely related to soliton annihilation and survival. In Fig. 1(d) of the main text, solitons annihilate in all the scans under strong thermal effects with $\xi=-1.2\times10^{5}$ (Ws)$^{-1}$. As a typical case, we show the evolution of $\Delta_{\text{eff}}$ under this strong thermal effects in Figs. \ref{FigS1Strong} (a), (b). We need to stop at a large laser detuning ($\Delta_{0}$=57) to induce soliton transition (choosing smaller $\Delta_{0}$ will only yield a chaotic comb). Solitons can form spontaneously after the laser scan ends. The intracavity power is 4.9 W within the soliton step region, which is not sufficient to maintain $\Delta_{\text{eff}}$ in the SSR; therefore the solitons annihilate.

Here, we give a simple estimation of the final power needed to support soliton survival in this case. When the cavity reaches a steady state, $d\delta_\Theta/dt=0$. This leads to ${{\delta }_{\Theta }}=\xi P{{\tau }_{0}}$. In order to have $\Delta_{\text{eff}}$ in the SSR, we should have,

\begin{equation}
9.2<2\xi P{{\tau }_{0}}/(\alpha+\kappa)+\Delta_{0}<17.9.
\label{eq:finaldetuning}
\end{equation}
Using $\tau_0$=100 ns, $\alpha+\kappa$=0.0037, $\xi=-1.2\times10^{5}$ (Ws)$^{-1}$, $\Delta_{0}$=57, the final power should be $6.03~\text{W}<P<7.37~\text{W}$. As 6 solitons correspond to a power of 3.39 W (Fig. 1(c) in the main text), this power requires the soliton number to be 11, 12 or 13, which is not accessed in the 50 scans.

In contrast, if thermal effects are weak ($\xi=-1.2\times10^{4}$ (Ws)$^{-1}$ in Fig. 1(e) of the main text), we can stop at a small laser detuning ($\Delta_{0}$=14) to induce the soliton transition (Figs. \ref{FigS1Strong}(c), (d)). The final intracavity power can be as low as 2.92 W to have $\Delta_{\text{eff}}$ in the SSR and support soliton survival. Based on Eq. \ref{eq:finaldetuning}, we find soliton number from 1 to 13 can all support soliton survival, with  $\xi=-1.2\times10^{4}$ (Ws)$^{-1}$, $\Delta_{0}$=14. Hence, the final intracavity power has minor effect on soliton annihilation and survival when $\xi$ is small.

\section{Experimental setup and method}
\begin{figure}[h]
\includegraphics[width=0.8\columnwidth]{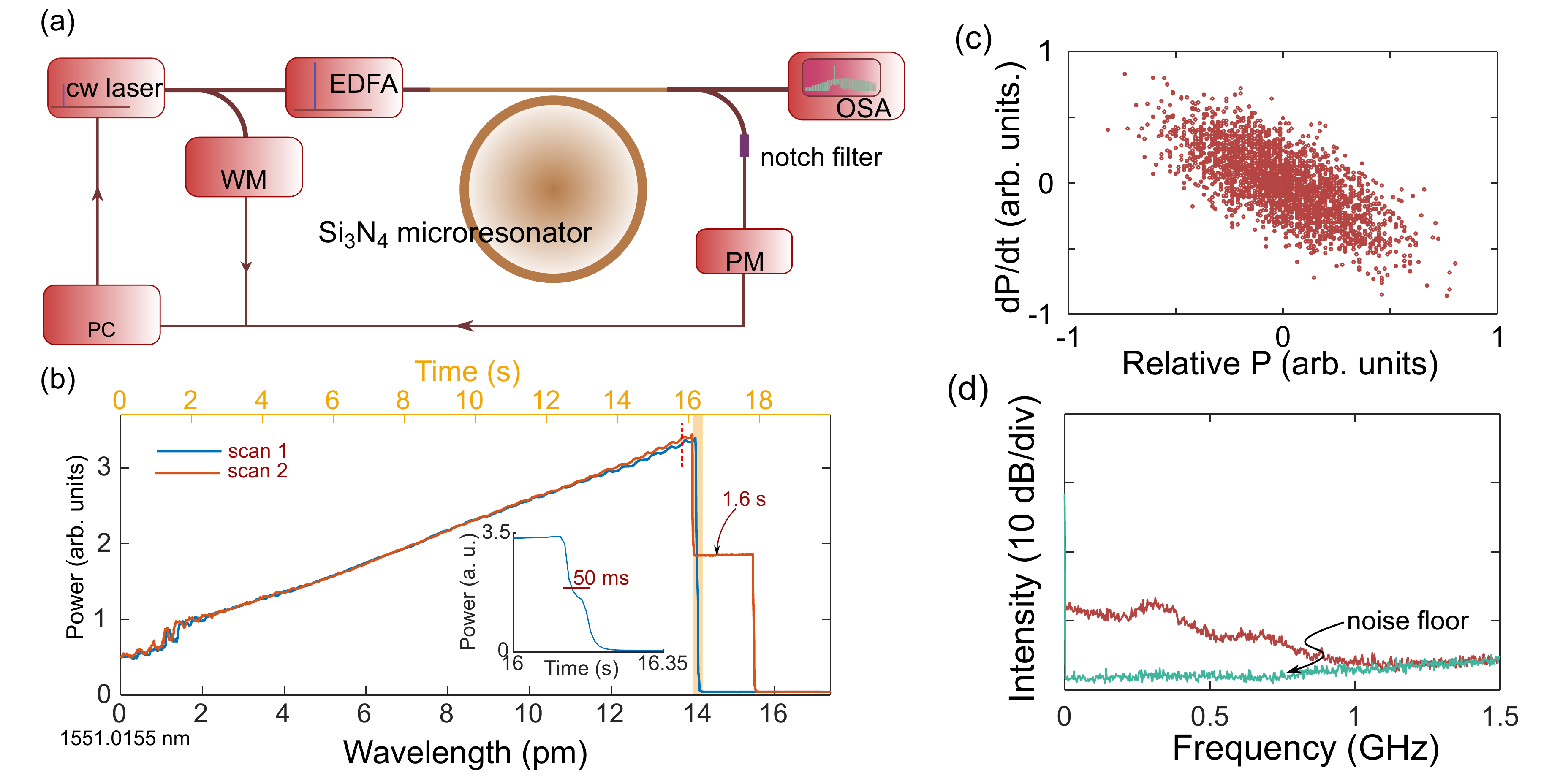}
\caption{(color online) (a) Experimental setup, in which a wavemeter is used for the calibration of the laser scan. WM: wave-meter, PM: power-meter, PC: computer. (b) Typical converted comb power traces when scanning across the resonance, showing distinct soliton step lengths. The inset shows the zoom in of the trace $scan~\emph{1}$ near the falling edge.  The phase diagram (c) and the corresponding rf spectrum (d) of the chaotic state, which corresponds to the state indicated by the dashed line in (b). }
\label{Fig1Setup}
\end{figure}

The experimental setup is shown in Fig. \ref{Fig1Setup}(a). The Si$_3$N$_4$ microresonator is pumped by an amplified cw laser. The converted comb power (after blocking the pump line by a notch filter) is measured during the comb generation. The tuning of the laser is also controlled by a computer; a wavemeter, calibrated by a cw laser stabilized to a femtosecond comb, is used to ensure the laser scan starts and ends at the predetermined wavelengths.

In DSG, we first scan the laser across the resonance to map out the soliton-step and then choose the stop wavelength based on the mapped soliton-step. In Fig. \ref{Fig1Setup}(b), we show two typical converted comb power traces. We can see distinct soliton-steps in different scans. The soliton-step length can be $>$100 ms in some cases. Here, we choose a wavelength near the middle of the soliton-step of $scan~\emph{1}$ (1551.0301 nm) as the stop wavelength of DSG.

For wavelengths before the falling edge of the soliton-step, the comb is in the chaotic state. As an example, we stop the laser at the wavelength indicated by the dashed line in Fig. \ref{Fig1Setup}(b) and measure the output power of the converted comb by a photodetector. The phase diagram (a plot showing the derivative of the power with respect to time against power) of the measured power trace and the corresponding rf spectrum are demonstrated in Figs. \ref{Fig1Setup}(c),(d), respectively. The phase diagram and the broadband intensity noise show chaos of the comb power.

\section{Direct soliton generation under step-tuning and shorter thermal response time}

\begin{figure}[t]
\includegraphics[width=0.9\columnwidth]{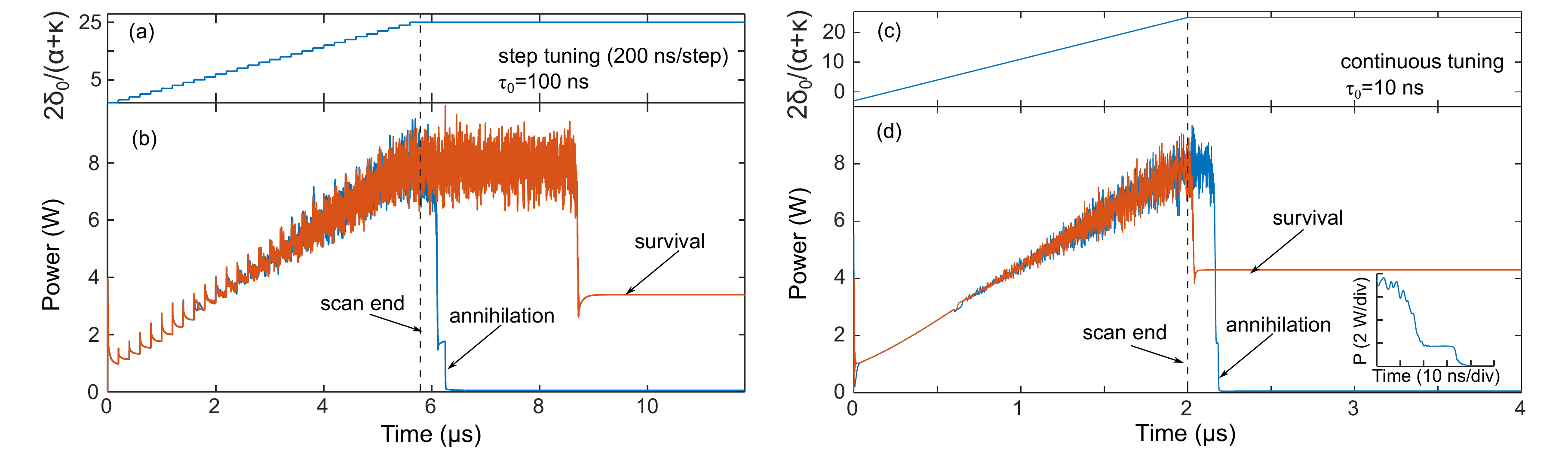}
\caption{(color online) (a) Tuning of the laser detuning $\Delta_0$ in steps from $-$3 to 25 with increment of 1, and holding at 25 for DSG. (b) The intracavity power evolution in slow time. The coexistence of soliton annihilation and survival and spontaneous soliton generation can still be found in the step tuning method. (c) Continuous tuning of the laser from $-$3 to 25 and holding at 25, with shorter thermal response $\tau_0$=10 ns. (d) Intracavity power change shows the coexistence of soliton annihilation and survival, as well as the spontaneous soliton generation. The inset is the zoom-in of the soliton step in the soliton annihilation case, showing a soliton step length $\sim$13 ns. }
\label{Fig1Speed}
\end{figure}

Limited by the computation ability, the laser tuning speed used in simulations is much higher than experiments. In experiments, every tuning of the laser takes $\sim$10 ms, much longer than the cavity lifetime. Therefore, the tuning of the laser can be regarded as step-like tuning. We show the simulated intracavity power change with step-like laser tuning in Fig. \ref{Fig1Speed}(a-b). The laser detuning $\Delta_0$ is tuned from $-$3 to 25 in a step-like way (the increment between steps is 1 and every step lasts for 200 ns, $\xi=-$4.5$\times$10$^{4}$ (Ws)$^{-1}$) and is held at 25, as shown in Fig. \ref{Fig1Speed}(a). The comb still exhibits the coexistence of soliton annihilation and survival with the step-like tuning, as shown in Fig. \ref{Fig1Speed}(b). Furthermore, the formation of solitons can take place after the end of the laser scan. Hence, the spontaneous soliton generation can also happen with stepped laser tuning.

Furthermore, we also tested DSG under shorter thermal response time. With a shorter thermal response time, the tuning of the laser is effectively slowed. Here, we decrease the thermal response time to 10 ns ($\xi$ increased to $-$4.5$\times$10$^{5}$ (Ws)$^{-1}$ accordingly to have same the thermal effects, see Sec. 1 and Eq. \ref{eq:finaldetuning} for the reason to increase $\xi$) and use the same tuning method as used in the main text (shown in Fig. \ref{Fig1Speed}(c)). The intracavity power change in Fig. \ref{Fig1Speed}(d) shows that coexistence of soliton annihilation and soliton survival still occurs with this effectively slower tuning. Also, the solitons can form spontaneously after the laser scan ends and annihilate (blue line in Fig. \ref{Fig1Speed}(d)). For the soliton annihilation case, the length of the soliton step is found to be $\sim$13 ns, much shorter than the length under $\tau_0$=100 ns ($>$100 ns, see Fig. \ref{Fig1Speed}(b) and Figs. 1(c), (d) in the main text). Since the backward-tuning method for soliton stabilization requires a long soliton-step, long thermal response time can be favored.

From the results in Fig. \ref{Fig1Speed}, despite the discrepancy in the laser tuning speed in the simulations and experiments, the general characteristics, e.g., coexistence of soliton annihilation and survival, and spontaneous soliton generation in DSG are captured.

\section{Direct soliton generation under higher pump power and larger detuning}

\begin{figure}[h]
\includegraphics[width=0.95\columnwidth]{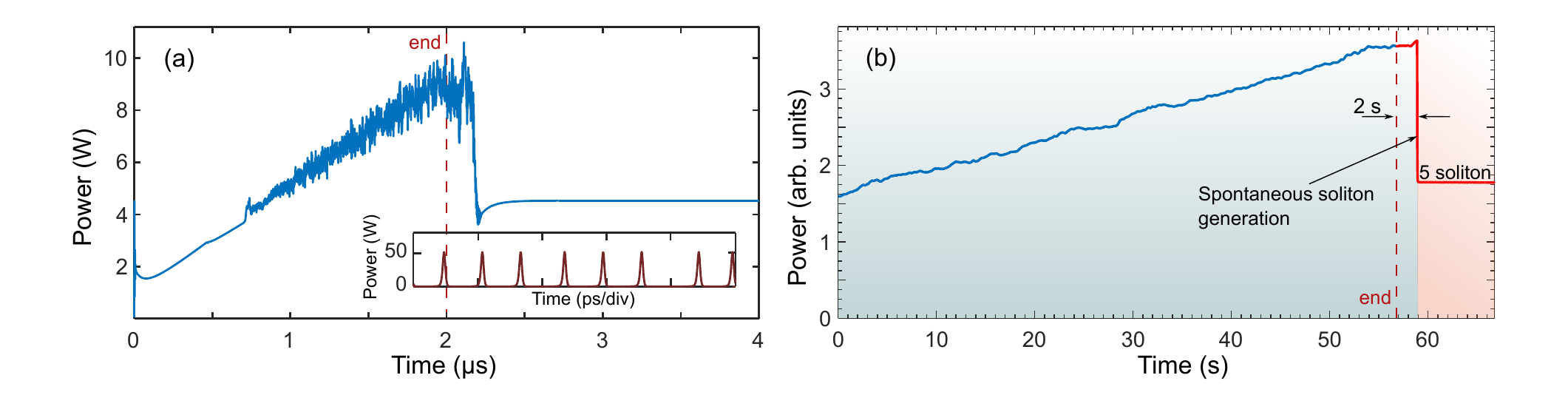}
\caption{(color online) (a) Simulated power trace for the case of soliton survival with larger pump power and $\Delta_0$ than in Figs. 1(b, c) of the main text. The inset shows the final comb consists of 8 solitons. (b) Experimentally measured converted comb power trace shows the spontaneous soliton generation, when stopping the laser wavelength at 1551.0323 nm under 300 mW pump power. The comb remains in the chaotic regime (blue-detuned to the hot cavity resonance) when laser scan ends and transits to the soliton state spontaneously after 2 s.  The red-dashed lines show the end of the laser scan.}
\label{Fig2large}
\end{figure}

The number of solitons generated is larger when using a higher pump power and larger detuning in simulations, similar to experimental results in Fig. 3(f) of the main text. For instance, if we increase $\kappa|E_{in}|^2$ to 0.13 mW, and tune $\Delta_0$  from $-$3 to 28.3 linearly in 0$\sim$2 $\mu$s and hold it at 28.3 in 2$\sim$4 $\mu$s, the soliton number is 8 for the case of soliton survival (inset of Fig. \ref{Fig2large}(a)), larger than the 6 solitons in Fig. 1(c). Moreover, the final intracavity power is increased to 4.5 W, compared to 3.4 W in Fig. 1(c) of the main text.

In Fig. 3(f) of the main text, solitons always form before the laser scan ends, as we use a large laser detuning. When choosing a small laser detuning (e.g., stop the laser at 1551.0323 nm), we can also observe the spontaneous soliton generation under the pump power of 300 mW. The corresponding power trace is shown in Fig. \ref{Fig2large}(b). The comb remains chaotic (pump inferred to be blue-detuned to the hot cavity resonance), when the laser scan ends. After holding the laser for 2 s, it transits to the soliton state spontaneously and survives in the soliton state.

\parskip 12pt